\documentclass[10pt,twocolumn,letterpaper]{article}

\usepackage[pagenumbers]{cvpr} 
\usepackage{amsmath,amsfonts}
\usepackage{algorithmic}
\usepackage{array}
\usepackage{textcomp}
\usepackage{url}
\usepackage{verbatim}
\usepackage{graphicx}

\usepackage{caption}
\usepackage{times,epsfig,graphicx,amsmath,amssymb,booktabs,tabulary,multirow,overpic,graphicx,xspace}
\usepackage{wrapfig}
\usepackage{enumitem}
\usepackage{adjustbox}
\usepackage{comment}
\usepackage{color}
\usepackage{tabu}
\usepackage{subcaption}

\usepackage{algorithm}
\usepackage{algorithmic}
\usepackage{listings}

\usepackage{soul}
\usepackage{dblfloatfix}
\usepackage{transparent}

\usepackage[dvipsnames]{xcolor}

\definecolor{cvprblue}{rgb}{0.21,0.49,0.74}
\usepackage[pagebackref,breaklinks,colorlinks,citecolor=cvprblue]{hyperref}

\usepackage{textpos}  
\usepackage[utf8]{inputenc}
\usepackage{amsthm}
\usepackage[switch]{lineno}

\title{FLARE: A Framework for Stellar Flare Forecasting using Stellar Physical Properties and Historical Records}

\author{
Bingke Zhu$^{1,}$\thanks{Equal contribution.}\quad
Xiaoxiao Wang$^{1,}$\footnotemark[1]\quad
Minghui Jia$^{2}$\quad
Yihan Tao$^{2}$\quad
Xiao Kong$^{2}$ \\
Ali Luo$^{2}$ \quad
Yingying Chen$^{1}$\quad
Ming Tang$^{1}$\quad
Jinqiao Wang$^{1,3}$ \\
$^{1}$~Foundation Model Research Center, Institute of Automation, \\ Chinese Academy of Sciences, Beijing, China \\ 
$^{2}$~CAS Key Laboratory of Optical Astronomy, National Astronomical Observatories, Beijing, China \\
$^{3}$~Zidong Taichu (Beijing) Technology Co., Ltd., Beijing, China \\  
{\tt\small \{bingke.zhu, xiaoxiao.wang, yingying.chen, tangm, jqwang\}@nlpr.ia.ac.cn,} \\
{\tt\small jiamh@bao.ac.cn, \{y.tao, kongx, lal\}@nao.cas.cn} \\[0.5pt]
}

\begin{document}
\maketitle
 
\begin{abstract}
    Stellar flare events are critical observational samples for astronomical research; however, recorded flare events remain limited. Stellar flare forecasting can provide additional flare event samples to support research efforts. Despite this potential, no specialized models for stellar flare forecasting have been proposed to date. In this paper, we present extensive experimental evidence demonstrating that both stellar physical properties and historical flare records are valuable inputs for flare forecasting tasks. 
    We then introduce FLARE (Forecasting Light-curve-based Astronomical Records via features Ensemble), the first-of-its-kind large model specifically designed for stellar flare forecasting. FLARE integrates stellar physical properties and historical flare records through a novel Soft Prompt Module and Residual Record Fusion Module. Our experiments on the publicly available Kepler light curve dataset demonstrate that FLARE achieves superior performance compared to other methods across all evaluation metrics. Finally, we validate the forecast capability of our model through a comprehensive case study.
\end{abstract}

\section{Introduction}

\begin{figure}[!t]
  \centering
  \includegraphics[width=\linewidth]{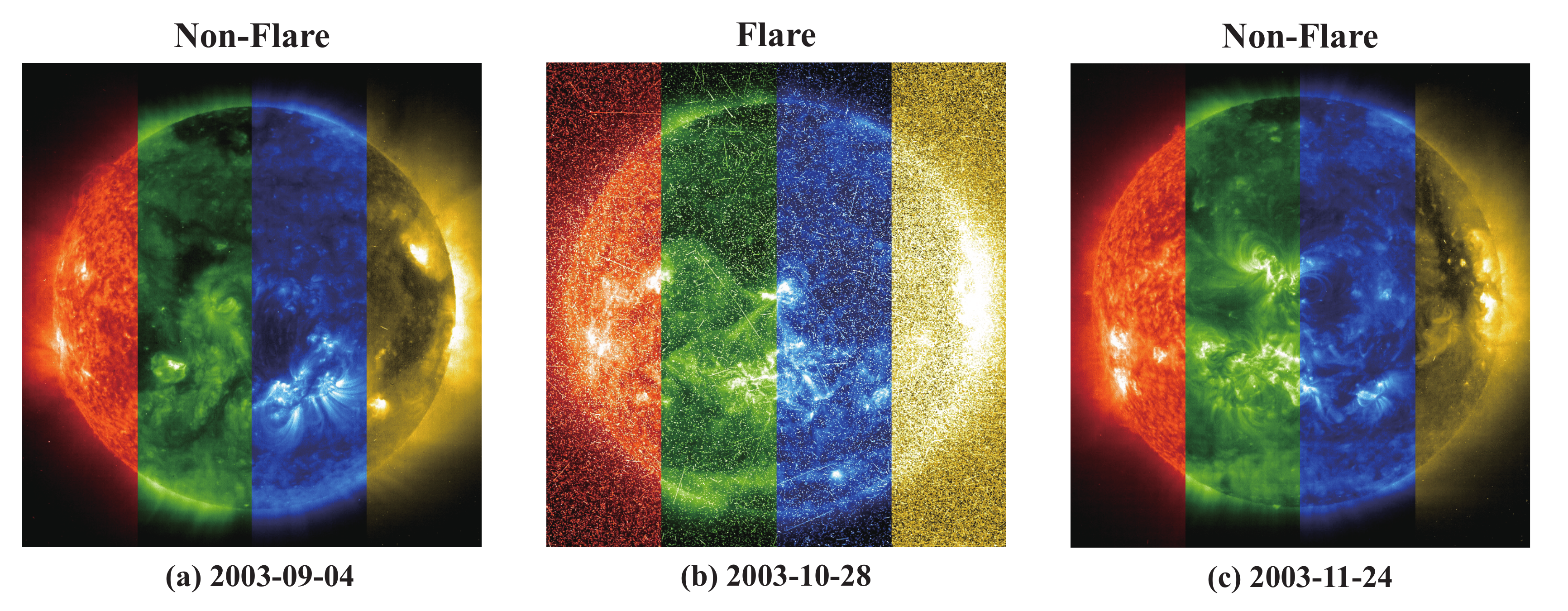}
  \caption{Star observations in multiple Extreme Ultraviolet (EUV) wavelengths before, during, and after a stellar flare.}
  \label{fig:sun_flare}
\end{figure}

Stellar flares are defined as the rapid release of magnetic field energy stored in a star's atmosphere, as illustrated in Figure~\ref{fig:sun_flare}. These phenomena are crucial for understanding stellar structure, evolution, and magnetic activity, as well as exploring potentially habitable exoplanets and extraterrestrial life~\cite{Yong_Research_2024}.
Flare records are currently obtained through continuous scanning of stars using survey telescopes in conjunction with manual analysis. Despite these efforts, the quantity of observed flare samples remains limited, rendering them inadequate for comprehensive research needs.
Consequently, forecasting stellar flare timing holds significant importance for astronomical studies. However, to date, there has been no published research addressing this area.

\begin{figure}[!t]
    \centering
    \includegraphics[width=\linewidth]{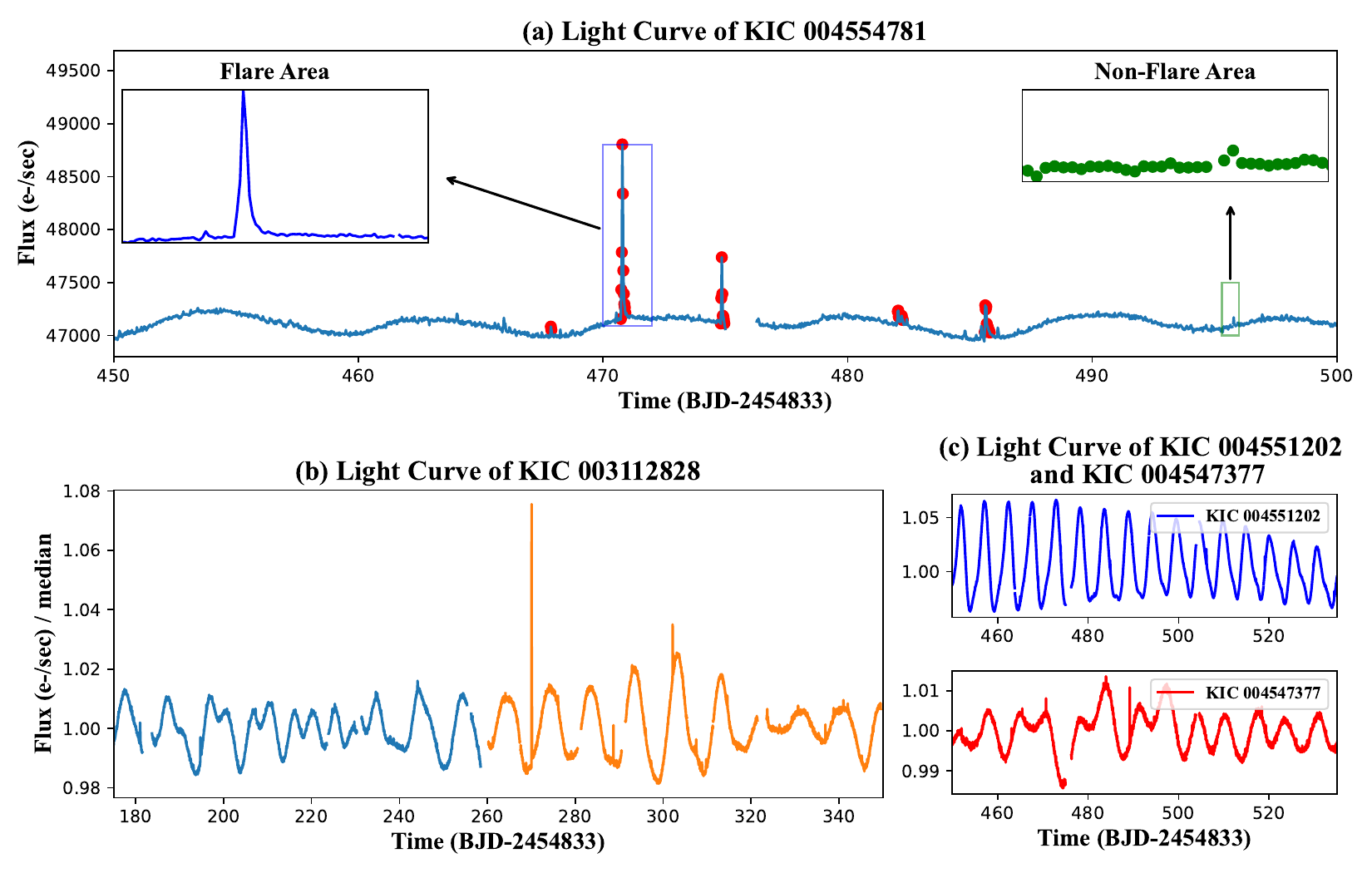}
    \caption{
    Light curves of several stars: (a) The flare region exhibits greater intensity in flux variations compared to non-flare regions. (b) The same star at different times displays distinct fluctuation patterns in its light curve. (c) Different stars during the same observation period display notable variations in their light curves, highlighting diversity across stellar systems.}
    \label{fig:flare_img}
\end{figure}

Solar flare prediction has garnered significant research attention~\cite{Deshmukh_Decreasing_2022,Abduallah_Operational_2023,Wen_Improving_2023}, but stellar flare forecasting presents distinct challenges compared to solar flare prediction.
Leveraging the solar proximity, researchers can easily obtain solar magnetograms and magnetic field parameters, facilitating accurate solar flare predictions. In contrast, stellar flare forecasting predominantly relies on light curves.
As depicted in Figure~\ref{fig:flare_img}, a light curve represents the chronological variation of a stellar luminosity, measured in flux using Julian Date as the time axis. This figure illustrates that light curves often have missing data points.
Additionally, two key characteristics emerge from the analysis:
(1) A single star exhibits varying trend patterns across different time periods (refer to Figure~\ref{fig:flare_img}(b)).
(2) The variation trends differ significantly among different stars (see Figure~\ref{fig:flare_img}(c)).
These complex variations in light curves pose challenges for flare forecasting. Observably, a flare event is characterized by a rapid flux increase followed by a gradual decline, resulting in sharp short-term flux changes. Conversely, non-flare regions do not display such characteristics (as shown in Figure~\ref{fig:flare_img}(a)).

The stellar flare forecasting task focuses on using light curves to predict whether a specified star will experience a flare within the next 24 hours. This can be viewed as a multi-task framework that combines both forecasting and classification objectives.
Deep learning methods~\cite{DBLP:conf/aaai/ZengCZ023,DBLP:journals/corr/abs-1803-01271,DBLP:conf/sigir/LaiCYL18,DBLP:conf/aaai/ZhouZPZLXZ21} have been widely applied to time series analysis with promising results in prior research. However, these methods still do not demonstrate superior feature extraction capabilities compared to pre-trained large multi-modal models in stellar flare forecasting. The intrinsic characteristics of stars, their varying evolutionary stages, and external factors such as other celestial bodies and interstellar dust can lead to diverse patterns in light curves. These external influences make achieving high accuracy in stellar flare forecasting using only light curves particularly challenging. As a result, additional data sources are required to improve the reliability of predictions.

In this paper, we introduce a novel task of forecasting stellar flare events. To address this challenge, we propose the FLARE framework (Forecasting Light-curve-based Astronomical Records via feature Ensemble). Through empirical analysis, we observe that stellar flares exhibit strong correlations with various stellar physical properties. Consequently, FLARE incorporates these stellar features as auxiliary inputs to enhance light curve feature extraction and improve forecasting performance. Furthermore, our investigation reveals that frequent historical flare events are positively correlated with the likelihood of future flares. To leverage this temporal dependency, FLARE integrates historical flare records as additional auxiliary features for enhanced prediction accuracy. We also introduce two novel components: the Soft Prompt Module, which combines stellar physical feature names and values to facilitate star-specific feature detection, and the Residual Record Fusion Module, designed to integrate light curves with historical flare records for improved model robustness. Finally, we employ a large multi-modal model fine-tuned using LoRA~\cite{DBLP:conf/iclr/HuSWALWWC22} to extract features from the outputs of these modules, thereby enabling accurate stellar flare forecasting.

The main contributions are as follows: (1) We present the first attempt at developing a method for stellar flare forecasting, addressing a previously unexplored challenge in astrophysics. (2) Through rigorous experimental analysis, we demonstrate that both stellar physical properties and historical flare records play significant roles in flare forecasting. Then, we propose a large-scale model called FLARE, which has shown remarkable effectiveness in enhancing accuracy. (3) Extensive experimental results validate the superior performance of FLARE compared to other approaches.
 
\section{Literature Review}

\subsection{Time Series Representation Learning}

Time series representation learning methods can be categorized based on the backbone into five groups: MLPs, RNNs, CNNs, GNNs, and Transformers. 

Inspired by the efficient performance of autoregressive models, 
MLPs 
such as DLinear~\cite{DBLP:conf/aaai/ZengCZ023} demonstrate excellent performance. However, these methods often require additional design to capture time-wise dependency effectively. 
RNNs~\cite{DBLP:conf/sigir/LaiCYL18,DBLP:conf/ijcai/QinSCCJC17} are naturally suitable for modeling sequential data, while they suffer from issues such as gradient vanishing due to recurrent structure and struggle to learn relationships between multivariate variables. 
CNNs, unlike RNNs, are less prone to gradient vanishing and excel at capturing the local patterns in time series. However, they often require stacking multiple convolutional layers to learn global futures, as seen in TCN~\cite{DBLP:journals/corr/abs-1803-01271}, which results in a significant training time cost. 
GNNs~\cite{DBLP:conf/kdd/WuPL0CZ20,DBLP:conf/iclr/LiYS018} abstract variables as nodes and establish edges between multivariate variables, learning spatial dependencies through GCN~\cite{DBLP:conf/iclr/KipfW17}. However, this approach relies on message passing to capture global features, and shows less scalable than Transformers. 
Leveraging the self-attention mechanism, Transformers are particularly adept at learning long-term temporal dependencies and complex multivariate correlations. 
Transformers can be categorized into three categories according to different types of tokenization. 
Point-wise methods~\cite{DBLP:conf/nips/WuXWL21,Woo_ETSformer_2022} learn correlations between time steps but become computationally expensive for long sequences.
Series-wise methods~\cite{DBLP:conf/iclr/LiuHZWWML24} pay attention to model multivariate dependencies by tokenization, but struggle with complex temporal patterns.
Patch-wise methods~\cite{DBLP:conf/iclr/NieNSK23,DBLP:conf/iclr/ZhangY23} adjust patch sizes for flexibility across different time series, making them more adaptable to different types of time series data.
These methods have demonstrated certain advantages in specific tasks, and our work also adopts a patch-based approach in light curve processing.

\subsection{Time Series Analysis based on PLMs} 

Among time series large models, aside from MOMENT~\cite{DBLP:conf/icml/GoswamiSCCLD24} and Chronos~\cite{Ansari_Chronos_2024} which are trained from scratch using big time series data, most approaches are adaptations of existing PLMs. According to modification methods, these approaches can be sorted into three types:

\textbf{Fine-tuning.} Studies like UniTime~\cite{DBLP:conf/www/LiuHLDLHZ24} and OFA~\cite{zhou2023one} unfreeze a portion of parameters, while others, including TEMPO~\cite{DBLP:conf/iclr/CaoJAPZY024} and LLM4TS~\cite{Chang_LLM4TS_2024} leverage Parameter-Efficient Fine-Tuning (PEFT) methods, such as LoRA~\cite{DBLP:conf/iclr/HuSWALWWC22}, adapt to new data by increasing trainable parameters without disrupting the existing structure of the large model.

\textbf{Alignment.} PLMs trained on text data need to be aligned with time series data in the same data space. Based on the object of modification, these approaches can be divided into two categories. The first approach fine-tunes PLMs with time series data to map the model parameters into time series data space, as seen in LLM4TS~\cite{Chang_LLM4TS_2024}. The other approach maps the time series vector into text space, as demonstrated in TIME-LLM~\cite{DBLP:conf/iclr/0005WMCZSCLLPW24}, which employs multi-head attention mechanisms to achieve mapping.

\textbf{Prompt-learning.} Studies such as UniTime~\cite{DBLP:conf/www/LiuHLDLHZ24} incorporate text prompts, while TEST~\cite{DBLP:conf/iclr/Sun0LH24} employs the combination of trainable vectors and textual token embeddings to improve performance.

Although these methods generally perform well on specific tasks, they struggle to simultaneously handle multi-task time series analysis and text feature extraction.

\section{Preliminaries}
\begin{table*}[bth]
    \centering
    \resizebox{\textwidth}{!}{
    \begin{tabular}{l||cc|cc|cc|cc}
        \toprule [1pt]
        ~&\multicolumn{2}{c}{\textbf{Light Curve}}&\multicolumn{2}{c}{\textbf{Light Curve + HFRs }}&\multicolumn{2}{c}{\textbf{Light Curve + SPPs}} &\multicolumn{2}{c}{\textbf{Light Curve + HFRs + SPPs}} \\
           
           \cmidrule(rl){2-3} \cmidrule(rl){4-5} \cmidrule(rl){6-7} \cmidrule(rl){8-9}
           
       Methods& Accuracy & F1 score &Accuracy & F1 score &Accuracy & F1 score  &Accuracy & F1 score \\
        
        \cmidrule(rl){1-9}

        PatchTST &61.92 &69.69 & 64.52 &70.461&64.46 &70.01 &68.19 & 72.26 \\
        iTransformer &49.99 &66.67 & 64.73 &68.46&65.46 &71.25& 66.86 & 71.53 \\
        Autoformer & 50.01 & 66.67 & 64.28 & 67.94 & 66.10 & 71.01 & 66.52 & 70.99\\
        Crossformer & 50.00 & 66.67 & 65.33 & 68.69 & 65.64 & 71.21 & 66.58 & 71.35\\
        ETSformer & 50.00 & 66.67 & 65.40 & 69.03 & 65.67 & 71.12 & 67.19 & 70.83 \\
         
        \bottomrule [1pt]
    \end{tabular}
    }
    \caption{Performance on the Kepler light curve dataset with and without the use of Historical Flare Records (HFRs) and Stellar Physical Properties (SPPs) separately, and each number represents the result of a single experimental run. (\%)}

\label{tab:Empirical_observe}
\end{table*}

\subsection{Problem Definition}

The stellar flare forecasting task involves predicting whether a flare will occur in the future for a given star based on its light curve observations, physical properties, and historical flare records. For a star $i$, its physical properties are represented as $\mathcal{P}^i = \{(k_m^i, v_m^i) | m = 0, 1, \ldots, M\}$, where $k_m^i$ and $v_m^i$ are the name and the value of the $m$-th physical property, respectively. Here, $k_m^i$ is a text, while $v_m^i \in \mathbb{R} \cup \{\varnothing\}$, where $\varnothing$ indicates a missing value.
The observed flux for star $i$ at timestamp $t$ is denoted as $x_t^i \in \mathbb{R} \cup \{\varnothing\}$. The historical flare records are represented by $\mathcal{R}^i = \{(t_{s_n}^i, t_{e_n}^i) | n = 0, 1, \ldots, N\}$, where $t_{s_n}^i$ and $t_{e_n}^i$ denote the start and end times of the $n$-th flare, respectively. A binary indicator variable is defined such that $y_t^i = 1$ if a flare occurs at timestamp $t$, \ie, if there exists a time $t$ satisfying $t_{s_n}^i < t < t_{e_n}^i$; otherwise, $y_t^i = 0$. Additionally, $y_{(t_0, t_1)}^i = 1$ indicates that at least one flare occurred within the interval $(t_0, t_1)$.
Let $K$ and $H$ denote the observation window length and forecast horizon, respectively. The observed light curve is represented as $\vec{x}_t^i = [x_{t-K}^i, \ldots, x_{t-1}^i]^\top \in \mathbb{R}^K$, and the historical flare records are denoted by $\vec{y}_t^i = [y_{t-K}^i, \ldots, y_{t-1}^i]^\top \in \mathbb{R}^K$. The forecast probability of a flare occurring between timestamps $t$ and $t' = t + H - 1$ is denoted by $\hat{y}_{(t, t')}^i \in [0, 1]$.
The forecasting task can be formalized as follows:
\begin{equation}
    \hat{y}_{(t, t')}^i = F(\vec{x}_t^i; \vec{y}_t^i; \mathcal{P}^i; \Phi),
\end{equation}
where $\Phi$ represents the model parameters of the model $F$.

\subsection{Experimental Observations} \label{sec:exper_obse}
In Table \ref{tab:Empirical_observe}, we present an experimental analysis to investigate how stellar physical properties and historical flare records influence the accuracy of flare forecasting using Kepler dataset.

Since stellar physical properties exhibit distinct characteristics compared to light curves, we first map these property values into a compatible dimensional space before concatenating them with the light curve data to derive forecast probabilities. For historical flare records, which share similar temporal characteristics with light curves, $\vec{y}_t^i$ and $\vec{x}_t^i$ are concatenated along the flux dimension prior to model input.

The experimental results reveal that both stellar physical properties and historical flare records contribute meaningfully to forecasting performance. Furthermore, their combined use yields superior predictive accuracy compared to using either type of data alone.
Interestingly, we observe that the marginal gain in performance diminishes after incorporating additional supplementary data. This finding suggests that stellar flare forecasting should not be approached solely as a time series prediction task based on light curves. Instead, leveraging both stellar physical properties and historical flare records represents an effective strategy for enhancing forecast accuracy and achieving a more comprehensive understanding of the underlying phenomena.

\section{Methodology}
\begin{figure*}
    \centering
    \includegraphics[width=1\linewidth]{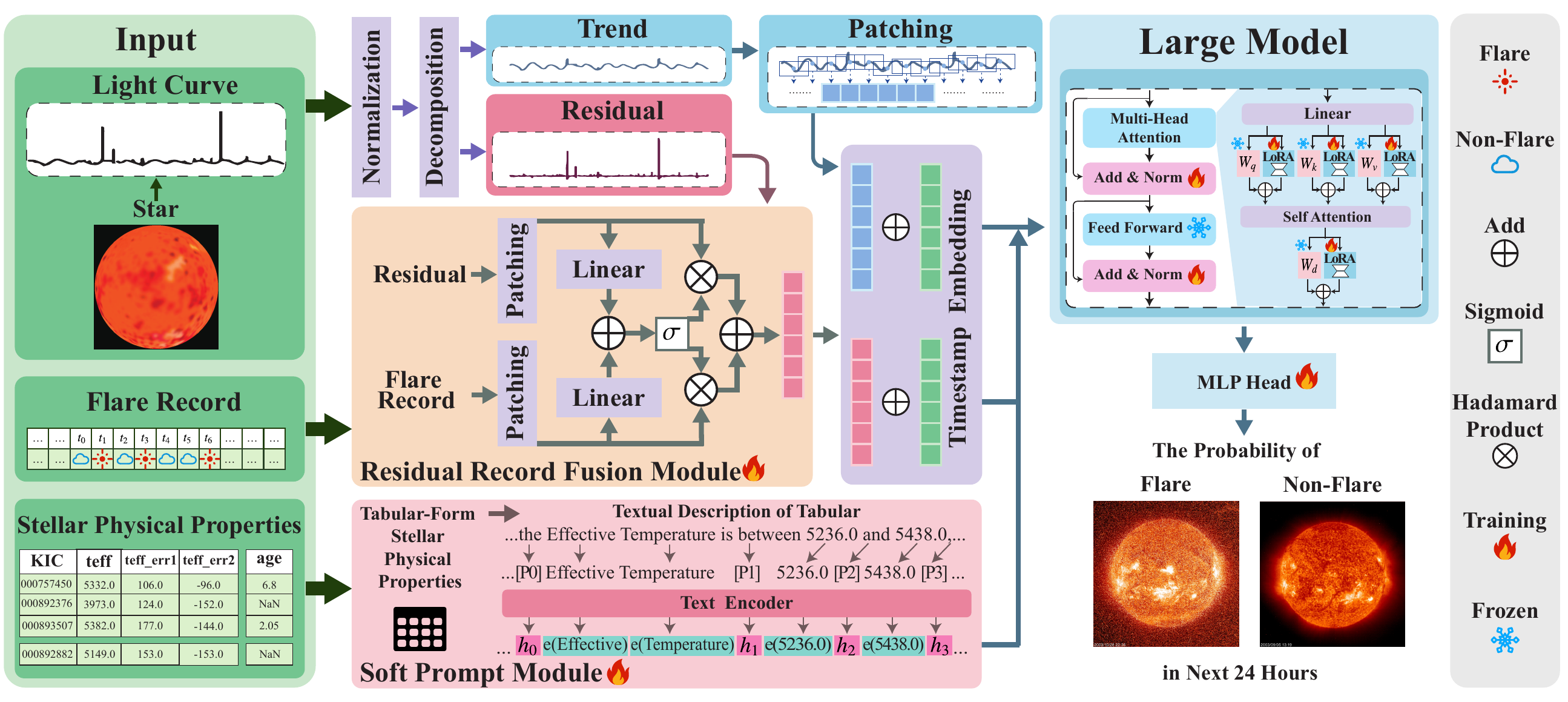}
    \caption{
    The overall framework of FLARE. First, the light curve is decomposed into trend and residual components, which are processed separately through patching and the Residual Record Fusion Module integrated with flare records. Timestamp embeddings are then appended to these processed components. Simultaneously, stellar physical properties are embedded using the Soft Prompt Module, generating a corresponding vector representation. The resulting vectors from both pathways are concatenated and passed through a large model. Finally, an MLP head processes the output to predict the probability of flare occurrence or non-flare conditions within the next 24 hours.}
    \label{fig:overview}
\end{figure*} 

In this section, we present our proposed model, FLARE. The overall architecture of the model is illustrated in Figure~\ref{fig:overview}, which comprises three key components. First, as detailed in Subsection~\ref{subsec:tsemb}, each light curve is decomposed into its trend and residual components, with historical flare records being integrated into the residual through a Residual Record Fusion Module to enhance robustness. Second, Subsection~\ref{subsec:prompt} introduces two prompt patterns based on tabular stellar physical properties and employs P-tuning~\cite{DBLP:conf/acl/LiuJFTDY022} to distinguish between different stars effectively. Finally, Subsection~\ref{subsec:llm} describes the fine-tuning of PLM to simultaneously process text and light curves. 

\subsection{Light Curve Embedding}\label{subsec:tsemb}


Typically, a flare only exists for a short period and is independent of the overall flux variation trend of the light curve. Furthermore, the periodic occurrence of flares only exists when the magnetic field of the star is stable. Given this phenomenon, separately handling the trend and the flare of the light curve could help eliminate mutual interference and improve light curve embedding. 
Even though the effectiveness of historical flare records has been verified in Subsection~\ref{sec:exper_obse}, fusing them with real flare could help improve robustness and reduce the misleading effect of false positive records.
However, not all stars exhibit clear periodic luminosity variations. Moreover, their periods vary significantly, ranging from shorter than one day to longer than the observation window. In such cases, real-time embedding can be beneficial.
Based on these considerations, we divide this subsection into three components: 
(1) normalization and decomposition, 
(2) trend processing and residual record fusion, 
and (3) timestamp embedding.  
As the light curve embedding process is uniform across stars, in the following text, we omit the superscript and use $x_t$ to represent the flux at timestamp $t$, and similarly for $y_t$.

\noindent\textbf{Normalization and Decomposition}. 
Due to the inherent limitations in the precision of the telescope, flux values exhibit substantial variations across different observation periods. This variation necessitates the normalization of the data by dividing the flux values by the median, in order to effectively mitigate the potential influence of systematic errors. 
Besides, frequent data omissions in light curves require us to perform a decomposition that distinguishes the overall flux trend from local abrupt flux variations, while minimizing the impact of missing values. 
Given that the time steps with missing data are usually non-consecutive, we employ a moving average to capture the trend of the light curve's variations, explicitly excluding the missing data from the computation to ensure that the trend is not unduly influenced by data gaps. This process can be represented as:


\begin{align}
    \hat{x}_j&=\frac{x_j}{\operatorname{median}(\vec{x}_t)},\\
    x_j^T&=\frac{1}{d_W}\sum_{j=t-\lfloor \frac{d_W}{2} \rfloor }^{t+\lfloor \frac{d_W}{2}   \rfloor }\hat{x}_jm_j, \\    
    m_j&= \begin{cases}
        0 & \text{if} ~ \hat{x}_j = \varnothing, \\
        1 & otherwise,
    \end{cases}    
\end{align}

\noindent where $\hat{x}_j$ is the normalized $x_j$, $m_j$ is an indicator for missing data, $d_W$ is the length of the sliding window, and $x_j^T$ represents the trend at timestamp $t$. $\vec{x}_t^T=[x^T_{t-K},\cdots, x^T_{t-1}] \in \mathbb{R}^K$ are utilized to represent the trend of the light curve, with the rest representing the residual $\vec{x}_t^R=[x^R_{t-K},\cdots, x^R_{t-1}] \in \mathbb{R}^K$, where $ x^R_j= \hat{x}_j -x_j^T, j=t-K,\cdots,t-1$.

\noindent\textbf{Trend Processing and Residual Record Fusion}. 
Since both the trend $\vec{x}_t^T$ and the residual $\vec{x}_t^R$ are univariate time series, the temporal context at each timestamp plays a crucial role in data embedding.
We generate patches for both the trend and the residual separately to obtain $X_t^T \in \mathbb{R}^{L\times P}$ and $X_t^R \in \mathbb{R}^{L\times P}$, where $L= \left\lfloor \frac{K - P}{S} \right\rfloor $ is the number of patches with length of $P$, and $S$ is the stride. 
The $X_t^T$ is passed through the MLP to obtain the $\hat{X}_t^T \in \mathbb{R}^{L\times d}$. 
For the residual, the gating mechanism is applied in conjunction with the flare record $\vec{y}_t$ to process $X_t^R$. This process can be represented as:
\begin{align}
    [\tilde{X}_t^R&;\tilde{Y}_t] =[X_t^R;Y_t] \tilde{W}+\tilde{b}, \\
    g=&\sigma(\tilde{X}_t^R W^{gR} +\tilde{Y}_t W^{gY} +b ^{g}), \\
    \hat{X}_t^R=&(g\odot \tilde{X}_t^R+ (1-g)\odot \tilde{Y}_t)W+b,
\end{align}
\noindent where $\tilde{W}\in \mathbb{R}^{P \times d}$, $\{W^{gR},W^{gY},W\} \subset \mathbb{R}^{d\times d}$, $\tilde{b}\in \mathbb{R}^{2L}$, $\{b^g,b\}\subset \mathbb{R}^{L}$. 
$Y_t$ is derived from $\vec{y}_t$ by generating patches. $\hat{X}_t^R \in \mathbb{R}^{L\times d}$ is the embedding of the residual. $\sigma$ denotes the sigmoid activation function and $\odot$ represents Hadamard product, respectively.

\noindent\textbf{Timestamp Embedding.} 
As stars exist beyond the solar system, the time formatting used on Earth is not suitable and is typically represented by Julian Date, which is a continuous floating-point number.
We extract the numerical values of each digit from the hundredths place to the ten-thousands place in a decimal manner, derive embeddings from these values, and use their sum as the timestamp embedding. 
Since the light curve has been divided into multiple patches, each patch requires a timestamp embedding. Given that the time intervals between adjacent time steps within the same patch are of fixed length, we employ the average timestamp embedding for the patch, and the collection of all patches is denoted as $E_t^T \in \mathbb{R}^{L\times d}$.

Following the three steps outlined above, the final time series embedding is obtained as $X_t^{TS}=[E_t^T+X_t^T;E_t^T+X_t^R] \in \mathbb{R}^{2L \times d}$.

\subsection{Prompt Design} \label{subsec:prompt}

\begin{figure}[ht]
    \centering
    \includegraphics[width=\linewidth]{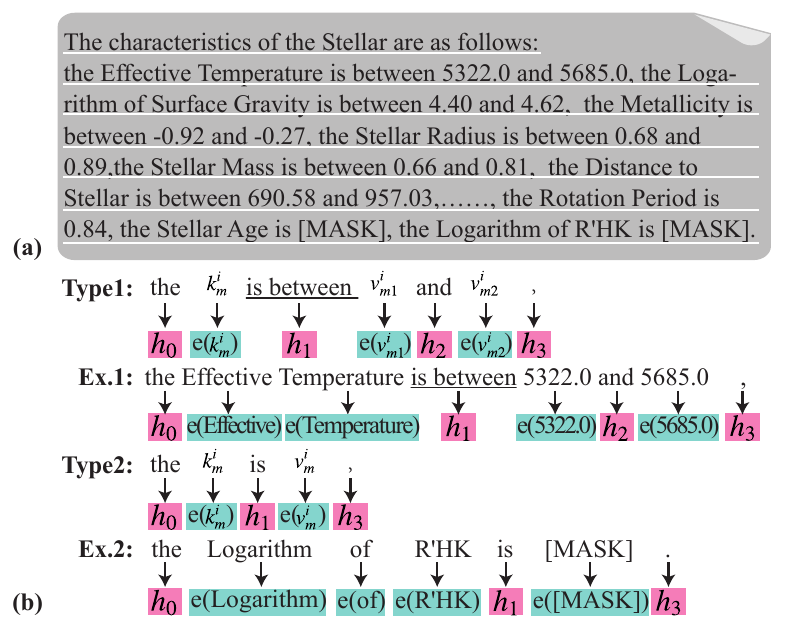}
    \caption{(a) A textual description of the star KIC 011924842's physical properties. (b) Two replacement pattens and examples.}
    \label{fig:prompt_tabular}
\end{figure}

\noindent\textbf{Motivation.} 
Stellar flares are a prominent manifestation of stellar activity. As summarized in Yong~\cite{Yong_Research_2024}'s study, factors such as stellar age, rotation speed, and stellar mass are correlated with flare frequency, which further supports the use of stellar physical properties for flare forecast task.
%
Stellar physical properties often exhibit frequent missing values and inconsistent numerical ranges. While interpolation and standardization can be used to address these issues separately, they may also introduce biases and lead to the loss of valuable physical information.
Given that the physical properties of stars are presented as tabular data, and that Hegselmann~\cite{DBLP:conf/aistats/HegselmannBLA0S23} has experimentally shown that combining column names with values leads to better performance than using only values, we organize the physical property values of stars along with their corresponding names into a textual structure. Furthermore, inspired by P-tuning~\cite{DBLP:conf/acl/LiuJFTDY022}, we design  the Soft Prompt Module to learn stellar physical properties for distinguishing stars.

\noindent\textbf{Soft Prompt Module.}
The textual description of stellar physical properties is shown in Figure~\ref{fig:prompt_tabular}(a), where any physical property is represented as a range of values or an exact value.
Inspired by P-tuning, which optimizes a small number of prompt embeddings and demonstrates good scalability while saving computational resources, we propose replacing part of the word vectors in the text with trainable parameters, as shown in Figure~\ref{fig:prompt_tabular}(b).
We design corresponding replacement patterns based on the types of the physical property. 
In both pattens, two vectors ($h_0,h_1$) are retained to represent the start and end of a the physical property description, with a vector $h_1$ separating the physical name from the value. 
Depending on the type of the physical property, an  additional feature separator vector $h_2$ will be inserted. 
All words except the replaced ones are embedded by the text encoder. 
Additionally, trainable embedding is utilized to represent the ID of the star.
Through this prompt design, the meaning of physical property names, as well as the physical significance of their numerical value ranges, are both preserved.
Finally, we use $X_p^i \in  \mathbb{R}^{S \times d}$ to represent the embedding of the physical properties of the star $i$, and $S$ is the number of tokens in the segmented textual descriptions.

\subsection{Pre-Trained Large Model Fine-tuning} \label{subsec:llm}
Zhou~\cite{zhou2023one} provides experimental evidence that training PLMs from scratch often hurts performance. 
However, freezing most of the parameters and only training a small subset can preserve the representational learning capability of PLMs. 
We freeze the majority of the parameters, particularly those in the multi-head attention mechanism and the feed forward layers, allowing the large model to fine-tune only the LayerNorm layers. To adapt to cross-modal inputs, we employ  Low-Rank Adaptation (LoRA)~\cite{DBLP:conf/iclr/HuSWALWWC22} to introduce trainable low-rank matrices to the multi-head mechanism, which allows effective learning of the correlation between the physical property text vectors $X_p^i \in  \mathbb{R}^{S \times d}$ and the light curve patches $X_t^{TS}$ while introducing only a small number of trainable parameters. 
The final embedding $Z_t^i=\text{PLM}(X_p^i;X_t^{TS}) \in \mathbb{R}^{2L\times d}$ is obtained at the end of this process.  

\subsection{Loss Function}
Since a portion of the samples are false positives, label smoothing is applied.
After computing the forecast flare probabilities with the MLP, we utilize the cross-entropy loss function incorporating label smoothing, which can be represented as:
\begin{align}
    \hat{y}^i_{(t,t')} =& \text{POOLING}(Z_t^i)W^c+b^c, \\
    \mathcal{L}_t^i =  &-[(1-\epsilon) y^i_{(t,t')} log(\hat{y}^i_{(t,t')})  \nonumber \\
       & +\epsilon (1-y^i_{(t,t')})log(1-\hat{y}^i_{(t,t')})], \\
    \mathcal{L} =&\frac{1}{\sum_{i=1}^{N_s} N_i} \sum_{i=1}^{N_s}\sum_{t=1}^{N_i} \mathcal{L}_t^i,
\end{align}
\noindent where $\epsilon$ is the smoothing coefficient of the label, $N_s$ denotes the number of stars, $N_i$ represents the number of samples for star $i$, $W^c \in \mathbb{R}^{1\times 2}$, $b^c \in \mathbb{R}^1$,  and POOLING refers to the operation of dimensionally reduction of $Z_t^i$.

\section{Experiments}

In this section, extensive experiments are conducted to evaluate the effectiveness of FLARE and the indispensability of each module, and  an analysis of several flare forecasting cases is presented. 

\subsection{Experimental Setup}

\noindent \textbf{Datasets.} Kepler mission~\cite{Borucki_KEPLER_2016} monitored the luminosity variations of over 150,000 stars from 2009 to 2018. 
For our study, we select high-precision light curves of 7,160 stars with flare events from 2009 to 2013, sampled every half-hour intervals, forming the Kepler light curve dataset. 
Each observation window consists of 512 data points, and the object is to forecast whether a flare event will occur within the next 24 hours, corresponding to 48 data points. The light curves of each star are split into training and test sets in a 4:1 ratio based on chronological order, and the flare rate of the test set is controlled at 50\% through random sampling.

\noindent \textbf{Baselines.} We compare the proposed method with five type baselines: 
(1) PLMs (MOMENT~\cite{DBLP:conf/icml/GoswamiSCCLD24}, Chronos~\cite{Ansari_Chronos_2024}, OFA~\cite{zhou2023one}, and UniTime~\cite{DBLP:conf/www/LiuHLDLHZ24})
(2) MLPs (Dlinear~\cite{DBLP:conf/aaai/ZengCZ023}, TiDE~\cite{DBLP:journals/tmlr/DasKLMSY23}, and FreTS~\cite{DBLP:conf/nips/YiZFWWHALCN23}), 
(2) RNNs (GRU~\cite{DBLP:conf/mwscas/DeyS17} and LSTNet~\cite{DBLP:conf/sigir/LaiCYL18}), 
(3) CNNs (MICN~\cite{DBLP:conf/iclr/Wang0HWCX23}, TCN~\cite{DBLP:journals/corr/abs-1803-01271}, and SCINet~\cite{DBLP:conf/nips/LiuZCXLM022}), 
(4) Transformers (PatchTST~\cite{DBLP:conf/iclr/NieNSK23}, iTransformer~\cite{DBLP:conf/iclr/LiuHZWWML24}, Autoformer~\cite{DBLP:conf/nips/WuXWL21}, Crossformer~\cite{DBLP:conf/iclr/ZhangY23}, ETSformer~\cite{Woo_ETSformer_2022}, and Informer~\cite{DBLP:conf/aaai/ZhouZPZLXZ21}).

\noindent \textbf{Evaluation Metrics.} To evaluate our forecasting model, we employ five evaluation metrics: AUC, Accuracy, Recall, F1 score and Precision. We prioritize high  Recall and Accuracy while keeping adequate Precision for accurate forecasting.

\noindent \textbf{Experiment Settings.} 
We use AdamW~\cite{DBLP:conf/iclr/LoshchilovH19} with a learning rate of 1e-5 as the optimizer, train for 200 epochs, and apply early stopping with a patience of 15. 
The PLMs, LSTNet, and TCN use publicly available code from their original papers, while the MLPs, Transformers, MICN, and SCINet use the code provided by TSLib\footnote{https://github.com/thuml/Time-Series-Library}. Both the text encoder and the PLM are BERT~\cite{DBLP:conf/naacl/DevlinCLT19}.

\begin{table*}[ht]
    
        \centering
    
        \resizebox{\textwidth}{!}{
        \begin{tabular}{l || cccc || l || cccc}
            \toprule [1pt]
            
            \multirow{2}{*}{ \textbf{Methods} } & \multicolumn{4}{c||}{\textbf{Light Curve }} &  \multirow{2}{*}{ \textbf{Methods} } & \multicolumn{4}{c}{\textbf{Light Curve + HFRs + SPPs}} \\
               \cmidrule(rl){2-5} \cmidrule(rl){7-10}
               
            & \textbf{Accuracy} & \textbf{F1 score} & \textbf{Recall} & \textbf{Precision} &  & \textbf{Accuracy} & \textbf{F1 score} & \textbf{Recall} & \textbf{Precision} \\
            
            \cmidrule(rl){1-10}
    
            MOMENT  & 62.65±0.40  & 70.02±0.04 & 87.25±0.77  & 58.48±0.40 &$\text{MOMENT}^*$ & 68.08±0.05 & 72.00±0.00  &82.09±0.16  & 64.13±0.09  \\
            
            Chronos  & 61.26±0.43  & 69.49±0.45 & 88.09±1.05  &57.34±0.42 &$\text{Chronos}^*$ & 65.01±0.31 & 70.88±0.08  &\textbf{85.08±0.82}  & 60.71±0.40   \\
            
            OFA & 49.99±0.00  & 66.66±0.00 & 99.99±0.00  &50.00±0.00&$\text{OFA}^*$ & 65.47±0.34 & 70.24±0.08  & 81.50±0.76  & 61.72±0.44   \\
            
            UniTime & 61.45±2.07  & 69.42±0.65 & 87.44±2.44  & 57.64±1.89 &$\text{UniTime}^*$ & 67.43±0.20 & 71.83±0.15  &83.06±0.90  & 63.28±0.35   \\
    
            \cmidrule(rl){1-10}
            DLinear &50.00±0.00  & 66.66±0.00 & 99.99±0.00  &50.00±0.00 &$\text{DLinear}^*$& 66.04±0.22 & 71.18±0.03  & 83.87±0.66 & 61.83±0.31  \\
            
            TiDE & 59.43±0.25  & 68.64±0.04 & 88.80±0.62  &55.95±0.22 &$\text{TiDE}^*$ & 66.04±0.73  & 71.14±0.02   & 83.71±1.72  &  61.89±0.97   \\
            FreTS & 49.99±0.00  & 66.66±0.00 & 99.99±0.00  &49.99±0.00&$\text{FreTS}^*$ & 65.64±0.48 & 70.25±0.04  & 81.12±1.10  & 61.96±0.66 \\
            
            \cmidrule(rl){1-10}
            
            MICN & 49.99±0.00  & 66.66±0.00 & 99.99±0.00  &50.00±0.00 &$\text{MICN}^*$ & 65.58±0.12  & 70.74±0.02  & 83.22±0.23  & 61.52±0.15   \\
            TCN & \texttt{OOT}  & \texttt{OOT} & \texttt{OOT}  & \texttt{OOT} &$\text{TCN}^*$ & \texttt{OOT}  & \texttt{OOT} & \texttt{OOT}  & \texttt{OOT}   \\
            SCINet & \texttt{OOT}  & \texttt{OOT} & \texttt{OOT}  & \texttt{OOT} &$\text{SCINet}^*$ & 67.38±0.61 & 71.88±0.12 &83.36±1.11 &63.19±0.78   \\
            \cmidrule(rl){1-10}
        
            GRU & 53.23±4.57 &  67.50±1.17 & 96.79±4.52  & 52.00±2.83 &$\text{GRU}^*$  & 66.54±0.63 & 71.14±0.04 &82.49±1.54  & 62.57±0.87  \\
            LSTNet & 60.05±0.58  & 68.69±0.06 & 87.67±1.16  & 56.48±0.51&$\text{LSTNet}^*$  & 67.84±0.45 & 71.55±0.15 &80.88±1.52  & 64.17±0.77  \\
            \cmidrule(rl){1-10}
    
            PatchTST & 62.09±0.19  & 69.74±0.04 & 87.40±0.33  &58.02±0.18 &$\text{PatchTST}^*$  & 68.40±0.29 & 72.29±0.05  & 82.48±0.08  & 64.38±0.46   \\
            
            iTransformer &50.00±0.00  & 66.66±0.00 & 99.99±0.00  &50.00±0.00 &$\text{iTransformer}^*$  & 66.30±0.53 & 70.81±0.09  & 81.74±1.18  & 62.47±0.72  \\
            
            Autoformer & 49.99±0.00  & 66.66±0.00 & 99.99±0.00  &50.00±0.00 &$\text{Autoformer}^*$ &63.62±0.80& 69.72±0.33  & 83.74±1.41  & 59.73±0.82   \\
            
            Crossformer &50.00±0.00  & 66.66±0.00 & 99.99±0.00  &50.00±0.00 &$\text{Crossformer}^*$  & 66.66±0.46 & 71.19±0.05  & 82.37±0.94  & 62.69±0.61  \\
            ETSformer &50.00±0.00  & 66.66±0.00 & 99.99±0.00  &50.00±0.00 &$\text{ETSformer}^*$  & 66.46±0.16 & 70.73±0.25  & 81.05±1.06  & 62.75±0.30 \\
    
             Informer  & 52.93±4.13  & 67.38±1.00 & 96.97±4.28  & 51.78±2.51 &$\text{Informer}^*$  & 65.42±0.30 & 70.45±0.03  &82.42±0.65  & 61.51±0.38   
             \\
            \cmidrule(rl){1-10}
    
            -  & -  & - & -  &- & FLARE &\textbf{71.65±0.35} & \textbf{74.11±0.02}  &81.11±1.04 & \textbf{68.22±0.72}   \\

            \bottomrule [1pt]
        \end{tabular}
        }
    
        \caption{Performance on the Kepler light curve dataset with and without the use of historical flare records (HFRs) and stellar physical properties (SPPs). A $\text{baseline}^*$ represents the baseline that combines features learned from the light curve and historical flare records with stellar physical properties, followed by forecasting through an MLP. \texttt{OOT} denotes that the running time exceeds 15 days. (\%) }
    
    \label{tab:result}
    \end{table*}

\begin{table*}[ht]
    \centering
    
  \begin{tabular}{l||ccccc}
        \toprule [1pt]
        Method   & AUC &Accuracy & F1 score & Recall & Precision \\
        \cmidrule(rl){1-6}
        FLARE w/o Soft Prompt Module     & 77.47     & 67.47 & 71.55 & \textbf{81.79} & 63.58 \\
        FLARE w/o Residual Record Fusion Module    & 79.22     & \textbf{71.80} & \textbf{74.11} & 80.73 & \textbf{68.50} \\
        FLARE w/o LoRA   & \underline{79.45} & 71.37 &   73.91   &\underline{81.11} & 67.89 \\
        FLARE  & \textbf{79.89}     & \underline{71.65} & \textbf{74.11}& \underline{81.11} & \underline{68.22} \\
        
        \bottomrule [1pt]
        
    \end{tabular}
    \caption{The ablation analysis of FLARE. Bold indicates the best, and underlining denotes the second-best. }
    \label{tab:ablation}
\end{table*}

\subsection{Performance Comparison}
We compare FLARE with various baselines, conducting at least three runs to compute the average performance, as depicted in Table~\ref{tab:result}. Here, FLARE clearly outperforms other methods and is the only one achieving an accuracy greater than 70\%.
The following are three key observations:

\begin{enumerate}
\item[(1)] Among the five types of baselines, PLMs and RNNs typically perform well when neither historical flare records nor stellar physical properties are employed. The high effectiveness of MOMENT and Chrones can be attributed to the knowledge gleaned from pre-trained large time-series models. We further analyze the subpar performance of OFA, which results from its simplistic approach to light curve processing. In contrast, UniTime, based on the same PLM, performs commendably. Additionally, the strong performance of RNNs is ascribed to the temporal characteristics of light curves.
\item[(2)] Among MLPs and Transformers, only TiDE and PatchTST show classification ability when restricted to using only light curves. An analysis of this phenomenon is presented. The robust performance of TiDE is credited to the Residual Block, which bolsters its resilience and enables it to manage a certain level of noisy samples in the dataset. Among Transformers, point-wise methods (\eg, Autoformer and ETSformer) perform poorly, presumably because flux values lack contextual information, rendering them inadequate for effective feature learning at each time step. Series-wise methods (\eg, iTransformer) have difficulty capturing complex temporal dependencies, while patch-wise methods, such as PatchTST, exhibit excellent performance. Although Crossformer, also a patch-wise method, shows poor metrics, our analysis indicates that this is due to the mismatch between the univariate light curve and the Cross-Dimension Attention Mechanism.
\item[(3)] The inclusion of historical flare records and stellar physical properties improves the performance of all baselines, with only minor metric differences. This result underscores the significance of historical flare records and stellar physical properties in stellar flare prediction.
\end{enumerate}

\subsection{Ablation Study}

\paragraph{Effectiveness of Each Module}
An ablation study is conducted to assess the effectiveness of each module within FLARE. The results are presented in Table~\ref{tab:ablation}, where modules are either removed or replaced individually. ``FLARE w/o Soft Prompt Module'' indicates replacing the vector generated by the Soft Prompt Module with a vector mapped from stellar physical property values. ``FLARE w/o Residual Record Fusion Module'' means the removal of this module, with historical flare records concatenated with the light curve along the flux dimension. ``FLARE w/o LoRA'' refers to the exclusion of LoRA during fine-tuning.

Our evaluation reveals that the absence of any of the three modules leads to a decline in certain performance metrics, while FLARE exhibits robust performance across all metrics. Notably, the omission of the Soft Prompt Module results in a significant performance drop. This finding aligns with the conclusions of Hegselmann~\cite{DBLP:conf/aistats/HegselmannBLA0S23}, which posits that textual headers contribute to the classification of tabular-form data. The removal of LoRA causes a slight decrease in all five metrics, demonstrating the utility of LoRA in fine-tuning. Although the performance change before and after removing the Residual Record Fusion Module is minimal, retaining it leads to a higher Recall, thereby highlighting the effectiveness of this module.

\paragraph{Ablation Studies on Text Encoder and PLM.}

\begin{table}[!t]
    \centering
    \resizebox{0.5\textwidth}{!}{
    \begin{tabular}{l||cccc}
        \toprule [1pt]
        \textbf{Text Encoder + PLM}   &\textbf{Accuracy} & \textbf{F1 score} & \textbf{Recall} & \textbf{Precision} \\
        \cmidrule(rl){1-5}
        GPT-2 + GPT-2 &69.79 & 72.48 &79.58 & 66.55 \\
        BERT + GPT-2 &70.58 & 73.19 &80.32 & 67.23 \\
        BERT + BERT (ours)&\textbf{71.51} & \textbf{74.08} &\textbf{81.42} & \textbf{67.95} \\

        \bottomrule [1pt]
    \end{tabular}
    }
    \caption{The ablation analysis of the text encoder and PLM in FLARE.  Bold indicates the best.}
    \label{tab:textencoder_PLM}
\end{table}

An ablation study of the text encoder and PLM within FLARE is presented in Table~\ref{tab:textencoder_PLM}. The analysis reveals that when both components are GPT-2, the performance deteriorates. Conversely, upon the introduction of BERT, the performance improves. This improvement can be ascribed to the high-quality text embeddings that BERT provides for stellar physical properties. Employing BERT for both the text encoder and PLM yields the optimal performance, thereby highlighting the significance of the text encoder and PLM in FLARE.

\paragraph{Performance using only Light Curves.}

\begin{table}[!t]
    \centering
    \resizebox{0.5\textwidth}{!}{
    \begin{tabular}{l||cccc}
        \toprule [1pt]
        \textbf{Methods}    &\textbf{Accuracy} & \textbf{F1 score} & \textbf{Recall} & \textbf{Precision} \\
        \cmidrule(rl){1-5}
        SolarFlareNet&50.51&66.79&99.49&50.27 \\
        TS2Vec &63.35&68.81&82.72&58.90 \\
        CRT &60.08&68.12&85.26&56.72 \\
        LPT&57.67&68.55&94.07&54.96 \\
        VS-Loss &58.53&68.49&90.12&55.23 \\
        HIVECOTEV2& 56.43 & 67.18& 92.83& 52.64\\
        TARNet & 59.29 & 57.30 & 59.29 & 61.42 \\
        SAnD  & 50.00 & 66.67 & 1.00 & 50.00 \\
        
        \bottomrule [1pt]
    \end{tabular}
    }
    \caption{Performance of baselines using only Light Curves.}
    \label{tab:onlyLC}
\end{table}

A performance comparison of models using only light curves is presented in Table~\ref{tab:onlyLC}. The baselines include SolarFlareNet~\cite{Abduallah_Operational_2023}, TS2Vec~\cite{DBLP:conf/aaai/YueWDYHTX22}, CRT~\cite{DBLP:journals/tnn/ZhangYGH24}, LPT~\cite{DBLP:conf/iclr/DongZYZ23}, VS-Loss~\cite{DBLP:conf/nips/KiniPOT21}, HIVECOTEV2~\cite{DBLP:journals/ml/MiddlehurstLFLB21}, TARNet~\cite{DBLP:conf/kdd/Chowdhury0S0H22}, and SAnD~\cite{DBLP:conf/aaai/SongRTS18}. The comparison demonstrates the limited effectiveness of these models when relying solely on light curves. The results further highlight the efficacy of FLARE in leveraging historical flare records and stellar physical properties for forecasting stellar flares.

\paragraph{Ablation Studies on Pre-trained Large Models.}

\begin{table}[!t]
        \centering
        \resizebox{0.5\textwidth}{!}{
        \begin{tabular}{l||c||cccc}
            \toprule [1pt]
            \textbf{Methods}   & \textbf{Data} &\textbf{Accuracy} & \textbf{F1 score} & \textbf{Recall} & \textbf{Precision} \\
            \cmidrule(rl){1-6}
            ViT & LC+SPPs & 66.00 & 70.70 & 82.00 & 82.00 \\
            \cmidrule(rl){1-6}
    
             \multirow{2}{*}{ Chronos-tiny } & LC &60.93	&69.25	&87.99	&57.09\\
             ~& LC+SPPs & 64.67& 71.17 & 87.24 & 60.11\\
            \cmidrule(rl){1-6}
    
            \multirow{2}{*}{ Chronos-mini } & LC & 60.66	&69.15	&88.18	&56.88\\
            ~ & LC+SPPs & 65.67& 71.19 & 84.83 & 61.33\\
            \cmidrule(rl){1-6}
    
            \multirow{4}{*}{ MOMENT-small }  & LC & 59.52	&69.09&90.46	&55.88\\
            ~ & LC+HFRs & 63.18	&69.82	&85.17	&59.15\\
            ~ & LC+SPPs & 67.54	&72.09	&83.84	&63.23\\
            ~ & LC+HFRs+SPPs & 67.48& 71.99 & 83.60 & 63.21\\
            \cmidrule(rl){1-6}
            
           \multirow{4}{*}{ MOMENT-large } & LC & 57.96 &68.99	&93.55	&54.65\\
            ~& LC+HFRs & 63.96&	69.44	&81.90	&60.27\\
         ~ & LC+SPPs &68.52	&72.08	&81.27	&64.76\\
            ~& LC+HFRs+SPPs & 68.32& 72.70 & 84.39 & 63.86\\
    
            \bottomrule [1pt]
        \end{tabular}
        }
        \caption{Performance of pre-trained large models.}
        \label{tab:pre_large}
\end{table}

A performance comparison of pre-trained large models is presented in Table~\ref{tab:pre_large}. The models included are ViT~\cite{DBLP:conf/iclr/DosovitskiyB0WZ21}, Chronos~\cite{Ansari_Chronos_2024}, and MOMENT~\cite{DBLP:conf/icml/GoswamiSCCLD24}. This comparison demonstrates the impact of stellar physical properties (SPPs) and historical flare records (HFRs) on the stellar flare forecasting task.

\subsection{Case Study}

\begin{figure}[!t]
    \centering
    \includegraphics[width=\linewidth]{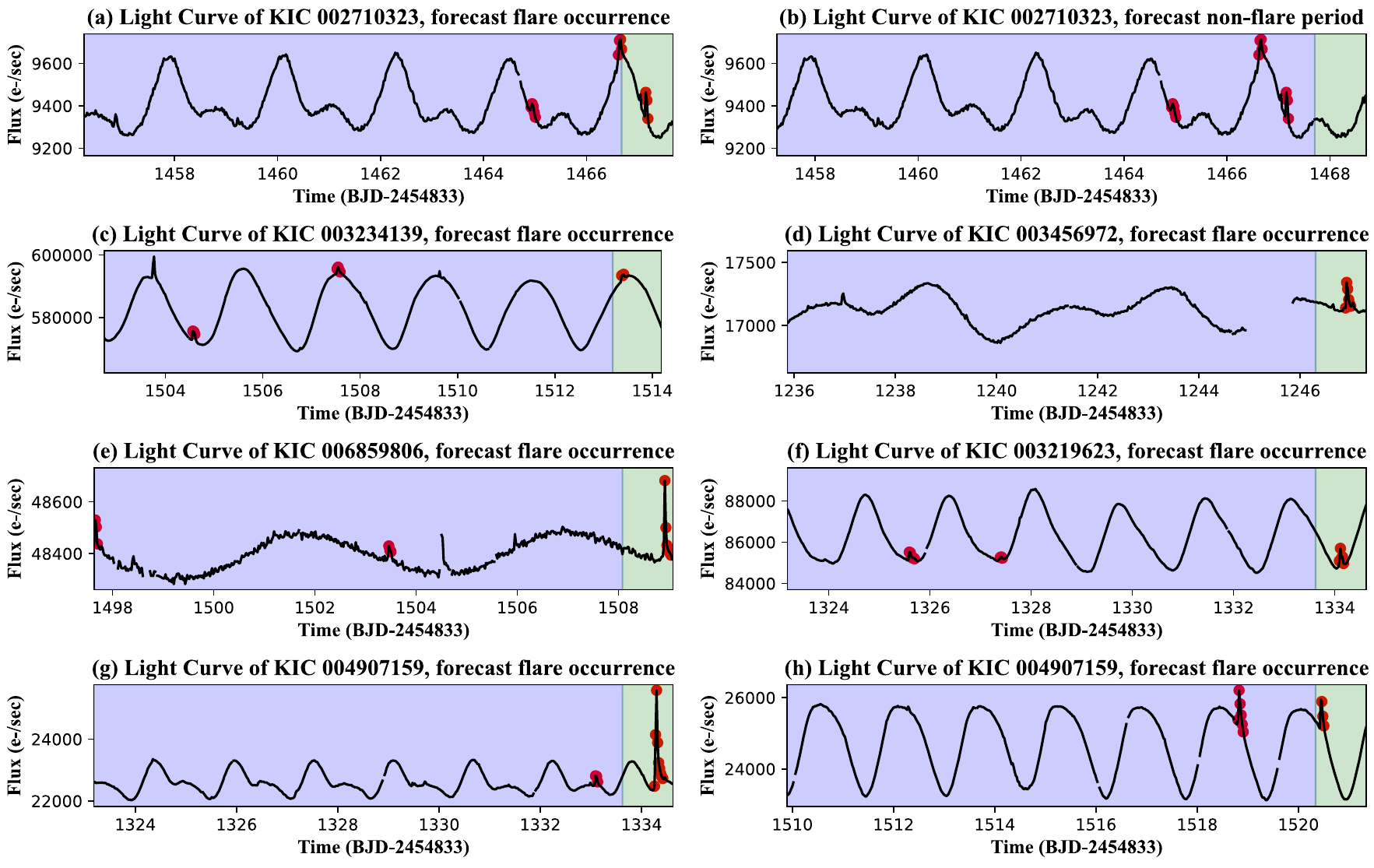}
    \caption{FLARE forecasts whether flares will occur in multiple samples. 
    The purple and green region represent the observation and the forecast area, and red dots mark the time steps that belong to the flares. 
    ``forecast flare occurrence" and ``forecast non-flare period" are used to represent the forecasting results of FLARE.
    } 
    \label{fig:case_study}
\end{figure}

To further elucidate the reasons underlying FLARE's flare forecasting capabilities, we conduct a case study to explore the working mechanism of FLARE. The forecast results of FLARE for selected samples are visualized in Figure~\ref{fig:case_study}. Evidently, FLARE can generate effective forecasts based on the observation area and can adapt to different stars and diverse flux variation patterns. Specifically, Figure~\ref{fig:case_study}(g) and Figure~\ref{fig:case_study}(h) demonstrate that FLARE can accurately predict flares on light curves with distinct flux variation patterns originating from the same star, thereby highlighting its robust forecasting ability.

\subsection{Statistical Observations}

\begin{figure}[!t]
\centering
    \includegraphics[width=\linewidth]{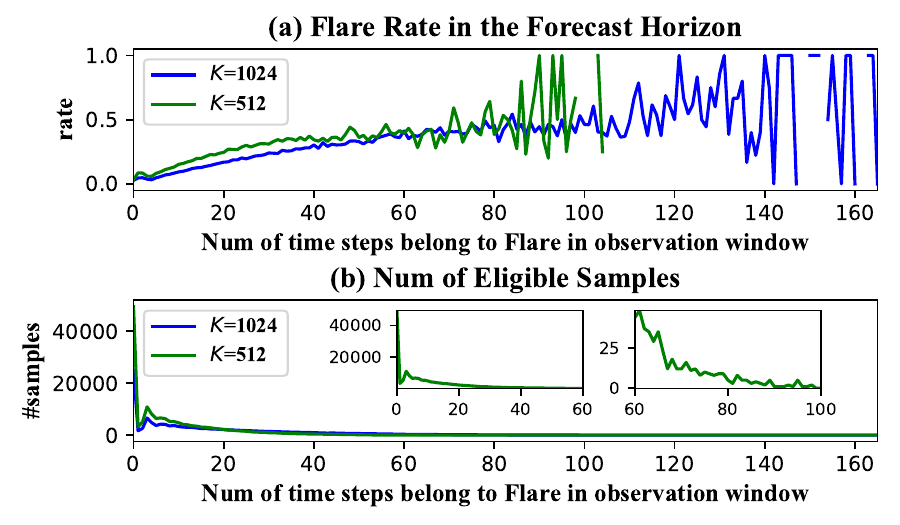}
    \caption{(a) Flare rates in the forecast horizon under different observation window lengths $K$. (b) The number of samples with certain number of time steps belonging to flare.}
    \label{fig:statis_observ}
\end{figure}

The analysis shown in Figure~\ref{fig:statis_observ} reveals that a positive correlation exists between the number of flare-related time steps within the observation region and the flare probability in the forecast horizon when the number of time steps corresponding to flares is relatively small. However, as the number of flare-associated time steps continues to increase, the quantity of eligible statistical samples decreases, resulting in a negligible correlation.

\subsection{Robust Analysis}

\begin{figure}[!t]
    \centering
    \includegraphics[width=\linewidth]{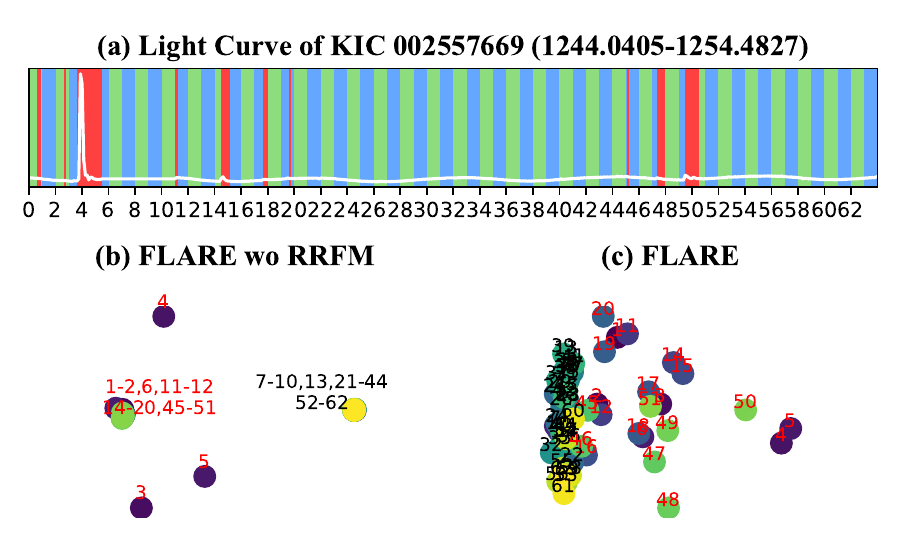}
    \caption{Robust analysis of a single case. (a) The light curve is partitioned into multiple patches, denoted by the green and blue intervals. The red area designates the flare region, which can be categorized into three types: \textbf{strong flare}, \textbf{weak flare}, and \textbf{suspected mislabeling}. Clearly, the 3-4th interval corresponds to a strong flare, the 14-16th and 48-52nd intervals fall within the weak flare region, while the remaining red areas are suspected of mislabeling. (b) and (c) present PCA visualizations of the features learned from each patch by FLARE, without and with the Residual Record Fusion Module (RRFM), respectively. The numbers indicate the sequential positions of patches, where red numbers signify patches containing at least one time step associated with a flare. }
    \label{fig:robust}
\end{figure}

To validate the robustness of the Residual Record Fusion Module (RRFM), we substitute it with residual features combined with embedded stellar flare records (FLARE w/o RRFM) and visualize the feature embeddings of each patch for both FLARE and FLARE w/o RRFM. As shown in Figure~\ref{fig:robust}(b), FLARE w/o RRFM can distinguish between strong flare and non-flare regions. However, it cannot differentiate between suspected mislabeling and weak flare regions. In Figure~\ref{fig:robust}(c), the dots, arranged from left to right, represent non-flare, suspected mislabeling, weak flare, and strong flare regions. The similarity between the features of the suspected mislabeling and non-flare regions highlights the enhanced robustness of FLARE against mislabeling.

\section{Conclusion}
In this paper, we demonstrate that both stellar physical properties and historical flare records are beneficial for forecasting stellar flares. Motivated by these findings, we propose the FLARE model, which incorporates two specialized modules: the Soft Prompt Module and the Residual Record Fusion Module. The Soft Prompt Module enables the model to differentiate between various star types, facilitating effective feature extraction tailored to each star's characteristics. Complementing this, the Residual Record Fusion Module enhances model robustness by integrating historical flare records with light curve residuals.
Our experiments on the Kepler light curve dataset underscore FLARE's superior performance compared to existing models. We expect that these empirical results will provide valuable insights for future advancements in stellar flare forecast research.

{
    \small
    \bibliographystyle{ieeenat_fullname}
    \bibliography{flare}
}


\end{document}